\documentstyle[aps,epsfig]{revtex}

\def\Journal#1#2#3#4{{#1} {\bf #2}, #3 (#4)}

\def\NPB{{\em Nucl. Phys.} B}
\def\NPA{{\em Nucl. Phys.} A}
\def\PLB{{\em Phys. Lett.} B}
\def\PRL{\em Phys. Rev. Lett.}

\def\PRC{{\em Phys. Rev.} C}

\def\PR{\em Phys. Rev.}

\def\FBS{{\em Few Body Systems Suppl.}}

\def\ply{{\rm Li}_2}

\begin{document}
\thispagestyle{empty}


\title {The Predictive Power of Effective Field Theory in NN Scattering: \newline ${}^3S_1$-${}^3D_1$ Mixing at Next-to Next-to Leading Order}

\author{Thomas D. Cohen and James M. Hansen}

\address{Department of 
Physics, University of~Maryland, College~Park, MD~20742-4111\\
DOE/ER/40762-191 \hspace*{.2in} UMPP \#00-012 }

\maketitle

\vspace{.25in}

\begin{abstract}
The convergence of the $Q$ expansion in the ${}^3S_1$-${}^3D_1$ mixing angle is investigated through the renormalization condition used to fit the free parameter encountered at NNLO.  Although the NNLO calculation describes the data well at momenta much less than $50$ MeV, a variation in the renormalization condition results in $100 \%$ expected uncertainties at momenta as low as $100$ MeV.  This uncertainty can be traced to the presence of large cancelations between terms entering at NNLO, which are found to be several times larger than the terms entering at NLO.  This implies a fundamental lack of predictive power at this order in the $Q$ expansion of $\epsilon_1$.  Given the small empirical size of $\epsilon_1$, this lack of predictive power is easily
understood in terms of simple power counting arguments.
   
 \vspace{.5in}
\end{abstract}

\newpage

\section{Introduction \label{intro}}

Effective field theory (EFT) descriptions of nuclear interactions are intended to
provide model independent predictions of observables using systematic expansions based
on a separation of
scales\cite{Weinberg1,Bira,Kap,vK97,Friara,Parka,cohena,Sa96,GPLa,DR,KSW,Kolck,KSW2,CGSSpol,Ccompt,SSpv,KSSWpv,SSWst,MehStew,Geg,SteFurn,CohHan1,CohHan2,CohHan3,Parkeft,KR,Epel,MSWsu,PBB,threebod,CRSnopi,PhilCoh,FMS,KS,BG}. In chiral perturbation theory
($\chi$PT) (the low energy effective field theory for hadronic interactions arising from
QCD) this is accomplished by implementing a power counting scheme which treats pionic
effects explicitly, while  heavier mesons  are integrated out leaving their effects to
be summarized in short distance parameters \cite{chiPT}.   Unlike the cases arising
in hadronic physics, the application of $\chi$PT to nuclear problems is not 
straightforward, as light scales exist in the problem (such as the scattering lengths)
which are formally small from the perspective of $\chi$PT, but are in fact unnaturally
large.

The original scheme proposed by Weinberg\cite{Weinberg1} to deal with the unnatural scales implements power
counting at the level of the Lagrangian, and then solves the full Schr\"odinger equation,
effectively iterating the interaction to all orders.  This scheme, also known
as $\Lambda$ counting, has enjoyed many phenomenological successes \cite{Bira,Parka,Parkeft,Epel,PhilCoh}. 
However, due to the nonperturbative nature of this approach, there is no
systematic power counting at the level of the amplitudes and observables. 

During the past two
years, there has been considerable interest in a new way to implement EFT for systems
of interacting nucleons originally developed by Kaplan, Savage and Wise (KSW)\cite{KSW}. 
The essence of this scheme is to count powers of a low mass
scale which is denoted by $Q$, with
 \begin{equation}
 m_\pi \sim   Q  \; \; \;
 1/a\sim  Q \; \; \;
 p \sim  Q \label{pc} \end{equation}
 where $a$ is the NN
scattering length, and $p$ is the relative momentum of the nucleon.

This $Q$ counting scheme is based on an expansion of  $Q/\Lambda$, where
$\Lambda$ is a  mass scale characteristic of the short distance physics.  Since
$m_\pi$, $1/a$ and $p$ are formally of the same order the scheme
includes $p a$, $m_\pi a$ and $p/m_\pi$ to all orders.  The scheme requires
that the lowest order NN contact interaction be iterated to all orders
building in the large scattering lengths.  All other interactions including all
pionic interactions are treated perturbatively.  This scheme has been used to
calculate numerous observables at next to leading order (NLO) with at least
modest success \cite{KSW2,CGSSpol,Ccompt,SSWst,MehStew}.
  The advantage of the scheme is its
rigorous power counting at the level of observables which allows for the {\it a
priori} estimate of the theoretical uncertainties in any regime where the $Q$
expansion is converging.  This ability to estimate higher order corrections is
one of the principal motivations for doing calculations with effective field
theory, rather than with potential models.  

It is also possible to develop an EFT where the pion is integrated out as heavy.  This pion-integrated-out theory is essentially equivalent to  zero range pseudo-potential   theory\cite{Bira}.  It contains an analog of $Q$ counting where the light scales are $p$ and $1/a$ only.  To avoid confusion, we will use the phrase, ``$Q$ counting,'' to refer to the full theory including explicit pions unless we explicitly indicate otherwise or it is clear from the context that we mean the pion integrated out case.  Apart from the ${}^3S_1$-${}^3D_1$ mixing angle, $\epsilon_1$, which is examined in this paper, calculations done with the 
pions-integrated-out EFT are often of comparable quality to calculations done to the same order in a theory with explicit pions\cite{KSW2,CohHan2,CohHan3,CRSnopi,BG}.  Surprisingly, for observables related to $S$-wave NN scattering, the 
pion-integrated-out theory fares significantly better than the theory with pions explicitly included at momentum less than $\sim \,100$MeV, even though the $p/\Lambda$ expansion has a radius of convergence limited by a cut at
 $p = i m_{\pi}/2$\cite{KSW2,CohHan2,CohHan3,CRSnopi}.

The fact that the pions-integrated-out theory often does as well, or better than
the theory with explicit pions raises questions about the convergence of the
chiral part of the $Q$ expansion, {\it i.e.}, the part based on the expansion
$m_\pi/\Lambda$\cite{CohHan2,CohHan3}.  The possible problems with the chiral $Q$ expansion are
particularly clear when one examines predictions for observables which are
chirally sensitive, in the sense that they are dominated by pion exchange
physics. Those observables which
vanish at some order in the $Q$ counting expansion for the pion-integrated-out
theory but which are finite at the same order in the theory with explicit
pions form an obvious class of chirally sensitive observables.  

An example of such a chirally sensitive quantity is the ``shape function'', ${\cal S}$,
introduced in Ref. \cite{CohHan3}.  It is defined as \begin{equation} {\cal
S}(p^2) \, = \, p \cot (\delta) \, - \, (-1/a \, + \, \frac{1}{2} r_e p^2)
\label{S} 
\end{equation} 
so that it removes the first two terms from the
effective range expansion and leaves the rest.  In $Q$ counting at NLO there
are two free parameters in S wave scattering which fix $a$ and $r_e$.  The
shape function removes sensitivity to the fit parameters by subtracting off the
first two terms in the effective range expansion.  In $Q$-counting
with explicit pions, ${\cal S}$ is order $(Q/\Lambda)^2$ due to the pion exchange
contributions; formally this can be seen in the inclusion of $p/m_\pi$ to all
orders.  In contrast, ${\cal S}$ is order $(Q/\Lambda)^3$ in the pion-integrated-out
theory.  Thus, in a regime where  $Q$ counting is valid, ${\cal S}$ is highly sensitive
to the pion dynamics.   Unfortunately, $Q$ counting fails to predict even the sign of
${\cal S}$ at momenta $\sim m_\pi$.  The failure of the theory to describe  ${\cal S}$  at NLO suggests that the chiral expansion part of the $Q$ counting may not be working
well. 

The conjecture  that the chiral part of $Q$ counting is converging poorly has been supported  by  recent calculations by Kaplan and Steele\cite{KS} using a toy potential including explicit one-pion-exchange with a realistic scattering length and effective range.  In this model, S-wave scattering  shows no sign of 
convergence at NNLO.   Of course, the model studied was
only a toy including only a one-pion-exchange potential.  It is logically
possible that higher-order calculations for the realistic nuclear
scattering problem which includes non-potential (``radiation pion'') effects\cite{MehStew}
will behave differently.  

It is important to check the convergence of the chiral part of $Q$ counting.
Ideally one should check convergence by doing higher-order calculations
for the realistic problem including radiation pion effects.
Unfortunately, at present a full NNLO calculation for s-wave scattering has not
been published.   However, the s-d mixing parameter (which involves the s-wave function at only NLO)  has been calculated at NNLO by Fleming, Mehan and Stewart
(FMS) \cite{FMS} including all radiation pion effects.  The s-d mixing parameter
should also be a good place to study the chiral part of $Q$ counting as it
is chirally sensitive in the sense defined above: it is zero at NLO in $Q$ counting for a theory with pions integrated out and nonzero in a theory with explicit pions.  FMS find extremely good agreement  between the NNLO calculation for 
the usual s-d mixing parameter, $\epsilon_{1}$\cite{epsdef},  and the $\epsilon_1$ extracted from the data (using the Nijmegen partial wave analysis (PWA)\cite{PWA}) for relative momenta up to 150 MeV \footnote{In FMS this mixing parameter is denoted by $\overline{\epsilon}_1$, and the mixing parameter in a different parameterization\cite{epsdef2} is called
 $\epsilon_1$}.  They conclude that this gives phenomenological support for the use of perturbative pions as is done in $Q$ counting.

In this paper, we 
investigate the dependence of $\epsilon_1$ at NNLO on the choice of the renormalization condition
 ({\it i.e.}, the method used to fit the one undetermined low energy constant).  The variations in $\epsilon_1$ resulting from different fits are found to be very
large; the variations are of order 100\% at $p$ of 100
MeV despite the fact that such variations only emerge at order $Q/\Lambda^3$ .  This sensitivity is a result of cancelations between terms entering at NNLO which are much larger than the total value of $\epsilon_1$ by typically an order of magnitude or more at momenta above 50 MeV.  Moreover one of the large and canceling terms is proportional to a free parameter and thus is entirely determined by the fit.  This strongly indicates that the large dependence upon the renormalization condition is a direct consequence of the lack of predictive power at this order in the $Q$ expansion.

\section{S-D Mixing  at NNLO}

 In this section the  NNLO calculation of $\epsilon_1$ done by Fleming, Mehan, and Stewart \cite{FMS} is discussed  briefly and relevant formulae
are presented. 

The PDS scheme of KSW\cite{KSW} is a renormalization scheme in which
the $Q$-counting is manifest.  The starting point is the effective Lagrangian:
\begin{eqnarray}  \label{Lpi}
{\cal L} &=& \frac{f^2}{8} {\rm Tr}\,( \partial^\mu\Sigma\: \partial_\mu 
\Sigma^\dagger )+\frac{f^2\omega}{4}\, {\rm Tr} (m_q \Sigma+m_q \Sigma^\dagger) 
+ N^\dagger \bigg( i D_0+\frac{\vec D^2}{2M} \bigg) N  \nonumber \\ 
&+& \frac{ig_A}2\, N^\dagger \sigma_i (\xi\partial_i\xi^\dagger -
\xi^\dagger\partial_i\xi) N -{C_0^{(^3S_1)}}
  {\cal O}_0^{(^3S_1)} +\frac{C_2^{(^3S_1)}}{8}{\cal O}_2^{(^3S_1)} 
  -{D_2^{(^3S_1)}} \omega {\rm Tr}(m^\xi ) {\cal O}_0^{(^3S_1)}  \nonumber\\[5pt]
 && -{C_2^{(SD)}} {\cal O}_2^{(SD)} + \ldots \,.
\end{eqnarray} 
where $f$ is the pion decay constant (approximately 132 MeV in this normalization), $\omega = m_\pi^2/(m_u + m_d)$ is related to the quark condensate and $f$,
  $\xi = \sqrt{\exp(i \sqrt{2} \vec{\tau}\cdot \vec{\pi}/f)} $ is related to the pion fields in the 
usual fashion in chiral perturbation theory, and the ${\cal O}$ are nucleon
contact interactions.  

The first step is to sum all graphs
containing only the contact interaction coupling $C_0$ to all orders which has the effect
of accommodating the unnaturally
long scattering length present in NN scattering.  As a result, the product $a
\, p$ is kept to all orders as required in $Q$ counting. The integral is divergent and is regulated using dimensional regularization with PDS. The resulting amplitude  is denoted by ${\cal A}^{(-1)}$.  All other interactions are included perturbatively.

Conventionally \cite{epsdef} the S matrix in the spin triplet channel is parameterized in the form 

\begin{equation}
S \, = \, \left( \begin{array}{l l} \cos(2 \epsilon_1) \, e^{ i (2 \delta_0)} \;&
\sin(2 \epsilon_1) \, e^{i (\delta_0 + \delta_2)}\;  \\ 
\sin(2 \epsilon_1) \, e^{i (\delta_0 + \delta_2)} &  \cos(2 \epsilon_1) \, e^{ i (
2\delta_0 )}
\end{array} \right )
\end{equation}  

\noindent where the above is taken to be the definition of $\epsilon_1$.  

At NLO $\delta_0$ depends on two parameters which are fit to the low energy scattering data, while $\epsilon_1$ and $\delta_2$ are both predictions.   At NNLO ({\it i.e.} order $(Q/\Lambda)^2$),  however, the $C_2^{\sc sd}$ coupling constant,  which summarizes short distance  s-d mixing also enters.  Thus at NNLO, there is one new
parameter which affects $\epsilon_1$ apart from those fit at NLO.

FMS derived an  expression for $\epsilon$ at order $(Q/\Lambda)^2$:
\begin{eqnarray}  
   \epsilon_1 \; &=&  \epsilon_1^{(1)} + \epsilon_1^{(2)} \\
   \epsilon_1^{(1)} &=&  \frac{M p}{4\pi}  \frac{ {\cal A}^{SD(0)} }
        {\Big[1+2\frac{ip M}{4\pi} {\cal A}^{(-1)}\Big]^{1/2} }     
     = \frac{M p}{4\pi} \ \Big| {\cal A}^{(-1)}\Big| \ 
        \frac{ {\cal A}^{{SD}(0)}}{ {\cal A}^{(-1)}}    \,, \\
   \epsilon_1^{(2)} &=& \frac{M p}{4\pi}  \frac{ {\cal A}^{{SD}(1)} }
        {\Big[1+2\frac{ip M}{4\pi} {\cal A}^{(-1)}\Big]^{1/2} } 
        -i \epsilon_1^{(1)}\Big[ \delta_0^{(1)} +  \delta_2^{(1)} \Big]  
     = \frac{M p}{4\pi} \ \Big| {\cal A}^{(-1)}\Big| \ {\rm Re}\Big[ 
        \frac{ {\cal A}^{{SD}(1)}}{ {\cal A}^{(-1)}}  \Big]  \,.  \label{eps2}
\end{eqnarray}

 where ${\cal A}^{(-1)}$ is the leading order contribution to the $S$ wave scattering amplitude given by:

\begin{eqnarray}
   {\cal A}^{(-1)} = -\frac{4\pi}{M} \frac1{\gamma+ip}\,,\qquad\quad 
 	\gamma=\frac{4\pi}{MC_0}+\mu_R  \,.
\end{eqnarray}

\noindent This amplitude contains a pole at $p = i \gamma$ representing the deuteron bound state.  Numerically, a binding energy of $2.22$MeV results in $\gamma = \sqrt{M B} = 45.7$MeV.    

\noindent The  s-d mixing amplitude involves one exchange of potential pions, and at order $Q^0$ is given by:  
\begin{eqnarray}  \label{Q0}
  {\cal A}^{SD(0)} &=& \sqrt{2} \frac{M g_A^2}{8\pi f^2}\: {\cal A}^{(-1)} 
       \bigg\{ m_\pi \, {\rm Re}[{\cal X(\alpha)}] - \frac{\gamma}{\alpha} 
       {\rm Im}[{\cal X(\alpha)}] \bigg\} \,, \\
   {\cal X}(\alpha) &=& -\frac{3}{4\alpha^2}-\frac{3i}{4\alpha}+\frac{i\alpha}2 + 
     i \bigg( \frac1{2\alpha} + \frac3{8\alpha^3} \bigg) \ln(1-2 i\alpha) \,, 
\label{chi}\end{eqnarray}
where
\begin{eqnarray}
  \alpha &\equiv & \frac{p}{m_\pi} \,.
\end{eqnarray}

The amplitude at order $Q$ involves the exchange of radiation pions, as well as potential pions.   The resulting s-d mixing amplitude at order $Q^2$ is given by:
  
\begin{eqnarray}  
 {\cal A}^{{SD}(1)} &=&  -{\cal A}^{(-1)}\ \zeta_6\: \alpha^2 - [{\cal A}^{(-1)}]^2 
   \sqrt{2}\: \frac{m_\pi^3 M g_A^2}{8\pi f^2} \: {\cal X}(\alpha)\ ( \zeta_1 \alpha^2 + 
   \zeta_2 ) \nonumber\\
   && + \sqrt{2} \frac{M m_\pi}{4\pi} \Big( \frac{g_A^2}{2 f^2} \Big)^2 \Bigg\{ 
     \frac{M m_\pi  {\cal A}^{(-1)}}{4\pi} \bigg[ 
    {\cal Z}(\alpha) + \frac{i}{2\alpha} \ln(1-2 i \alpha) {\cal X}(\alpha) \bigg] \nonumber \nonumber\\
   && \qquad\qquad\qquad\qquad\quad  - \bigg[ \frac{M m_\pi  {\cal A}^{(-1)}}{4\pi}
          \bigg]^2 \ln(1-2i\alpha) {\cal X}(\alpha) + {3 \over 2}{\cal Y}(\alpha) + 
         {\cal X}(\alpha)  \Bigg\} \,, \label{asd1} 
\end{eqnarray}
\noindent where ${\cal Y}(\alpha)$ and ${\cal Z}(\alpha)$ are given by: 
\begin{eqnarray}
{\cal Y}(\alpha) &=& -\frac25 + \frac3{10\alpha^2} + \bigg(\frac{3}{8\alpha^5} + 
      \frac{5}{4\alpha^3} - \frac{2\alpha}{5} \bigg) \tan^{-1}(\alpha)  
      -\bigg(\frac{3}{8\alpha^5} +\frac{5}{4\alpha^3} \bigg) \tan^{-1}(2\alpha) \nonumber \\*
   && + \frac{(15-4\alpha^2)}{80\alpha^6} \ln(1+\alpha^2)  
     - \frac{(3+16\alpha^2+16\alpha^4)}{32\alpha^7} {\rm Im}\,\bigg[  \ply\Big( 
      \frac{2\alpha^2+i\alpha}{1+4\alpha^2} \Big) +  \ply( -2\alpha^2-i\alpha )
      \bigg]  \nonumber \\*
   && + i \bigg[ \frac{3}{8\alpha^3} +\frac{1}{2\alpha} -\frac{\alpha}{2} - 
      \frac{(3+10\alpha^2)}{16\alpha^5}  \ln(1+4\alpha^2)
      +\frac{(3+16\alpha^2+16\alpha^4)}{128\alpha^7} \ln^2(1+4\alpha^2) \bigg]\,, \\ 
{\cal Z}(\alpha) &=& -\frac{7}{40} +\frac{9i}{16\alpha^3}+\frac{21}{40\alpha^2} +
  \frac{3i}{40\alpha}-\frac{3i\alpha}{5} + \frac{29\alpha^2}{200} + 
  \Big(\frac{3\alpha^2}{5} -\frac{9}{16\alpha^4}-\frac{15}{8\alpha^2} \Big) \ln{2} \nonumber \\*
 && + \frac{3\,(16\alpha^7-50\alpha^3-4i\alpha^2-15\alpha+15i)}{80\alpha^5} 
    \ln(1-i\alpha) \nonumber\\*
 && + \frac{(-9\,i + 27\,\alpha  - 24\,i\,{{\alpha }^2} + 
    78\,{{\alpha }^3} - 16\,{{\alpha }^5})}{32\alpha^5} \ln(1-2i\alpha) \nonumber \\*
 && - \frac{(9+48\alpha^2+48\alpha^4)}{64\alpha^6} \bigg[ \frac32 \ln^2(1-2i\alpha)
    + 2 \ply(-1+2i\alpha) + \ply \Big(\frac{1+2i\alpha}{-1+2i\alpha}\Big) +\frac{\pi^2}4
    \bigg]   \,.
\end{eqnarray}

$\zeta_1$, $\zeta_2$, and $\zeta_6$ are $\mu_R$ independent dimensionless 
combinations of coupling constants:
\begin{eqnarray} 
  \zeta_1 &=& {C_2(\mu_R) \over C_0(\mu_R)^2 } \,, \qquad
  \zeta_2 = {D_2(\mu_R) \over C_0(\mu_R)^2 } + {C_0^{(0)}(\mu_R) \over m_\pi^2
   C_0(\mu_R)^2} - {g_A^2 \over 2f^2} \bigg( \frac{M}{4\pi} \bigg)^2 \bigg[ \frac12 
    \ln\Big(\frac{\mu_R^2}{m_\pi^2}\Big) + {\mu_R^2 -\gamma^2 \over m_\pi^2} 
    \bigg]\,, \nonumber\\[10pt] 
 \zeta_6 &=& { m_\pi^2 C_2^{(SD)}(\mu_R) \over C_0(\mu_R)} + { 3\sqrt{2} \over 10}
    \bigg({M m_\pi g_A^2 \over 8\pi f^2 } \bigg)^2 \ln\Big({\mu_R^2 \over m_\pi^2}
    \Big)\,
\label{zetas}
\end{eqnarray}

The coefficients $\zeta_1$ and $\zeta_2$ are fit from the NLO
 s-wave phase shifts.   There is a certain amount of art in the
fitting of these coefficients. Numerically Mehan and Stewart find
$\zeta_1 = 0.2345$ and $\zeta_2 = -0.1038$ in fits to the Nijmegen partial wave analysis\cite{PWA} (PWA) heavily weighted
toward low momenta \cite{FMS}.    Note that although the preceding
amplitudes are order $(Q/\Lambda)^1$, from Eq.~(\ref{eps2}) they give rise
to a contribution to the mixing angle, $\epsilon_1$ at order $(Q/\Lambda)^2$.

\section{Sensitivity at NNLO}
   
The calculation of the ${}^3S_1$ - ${}^3D_1$ mixing angle, $\epsilon_1$, contains a single free parameter, $C_2^{\sc sd}$, at NNLO (apart from those fit for s-wave scattering at NLO).  This parameter contains all information about short distance mixing entering at this order, and appears in Eqs.~(\ref{asd1}) and (\ref{zetas}) as $\zeta_6$.  The linear combinations of Eq. (\ref{zetas}) explicitly absorb all $\mu_{\sc r}$ dependence into the undetermined short distance coupling constants. 
In essence, the fits of $\zeta_{1,2,3}$ are the renormalization conditions
 relating the parameters in the theory to physical quantities. 
In principle, one has a certain amount of freedom in choosing precisely
which physical quantity is fit.  Of course, to the extent the
$Q$ expansion is well converged to the correct result, one expects that the
dependence of the final prediction on the particular  choice of renormalization condition ({\it i.e.}, the particular quantity to which one fits)  will be confined to higher-order terms which are numerically small.  Thus, one way to 
test the sensitivity to higher order terms is by testing the sensitivity
to imposing different renormalization conditions.

\begin{figure}[t]
\epsfig{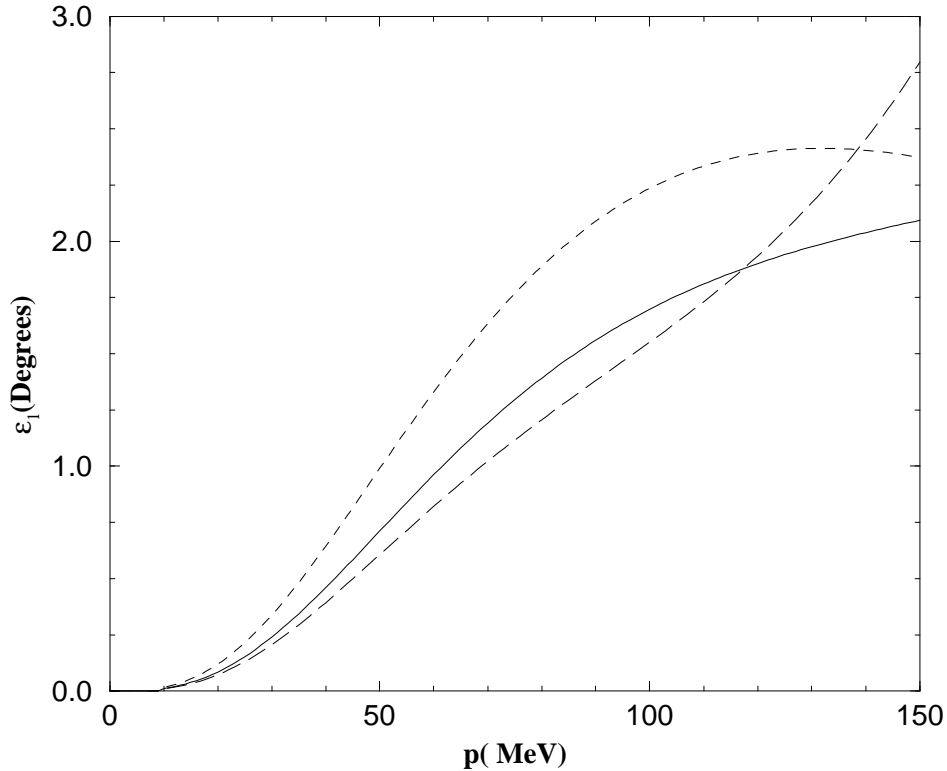}
\caption{Results of $Q$-counting calculations of $\epsilon_1$.  The solid line is the multi-energy Nijmegan partial wave analysis.  The short dashed curve is the leading order prediction in $Q$-counting.  The long dashed curve is the full NNLO calculation.}
\label{FMSfig}
\end{figure}

As noted above, $\zeta_{1,2}$ can be fit from $S$ wave scattering data at NLO.
Thus, the NNLO calculation of $\epsilon_1$ depends on a single free parameter
$\zeta_6$ associated with the s-d mixing counterterm.
 The method used in \cite{FMS} to fit $\zeta_6$ is via a global fit 
to the Nijmegen partial wave analysis of the scattering data, 
weighted towards low momenta; this results in $\zeta_6= 0.385$.   
The result of this fit for momenta up to $150$ MeV is given in Fig.~\ref{FMSfig}.
FMS conclude that the high quality of this fit
supports the use of perturbative pions as is required in the $Q$ counting scheme.

 In this paper we  investigate the sensitivity of the predicted
 s-d mixing angle to the choice
of renormalization condition in order to assess the contribution of higher-order terms.  We exploit 
 the fact that there is only one free parameter at this order.  Thus any
fitting procedure must be equivalent to fixing the predicted value of the 
mixing angle at some particular fixed value of momentum.  For example,
although the fit of FMS was a weighted global fit, the
 predictions  would have been the same had the renormalization condition 
for
 $\zeta_6$ been that the NNLO prediction for $\epsilon_1$ agree with the Nijmegen
PWA value at a momentum of approximately $111$ MeV.  Thus an efficient way to
explore different valid choices for renormalization conditions is to 
 construct fits which ensure that the theory agrees with the data at different fixed choices of momentum $p_{\sc r}$.

\begin{figure}[t]
\epsfig{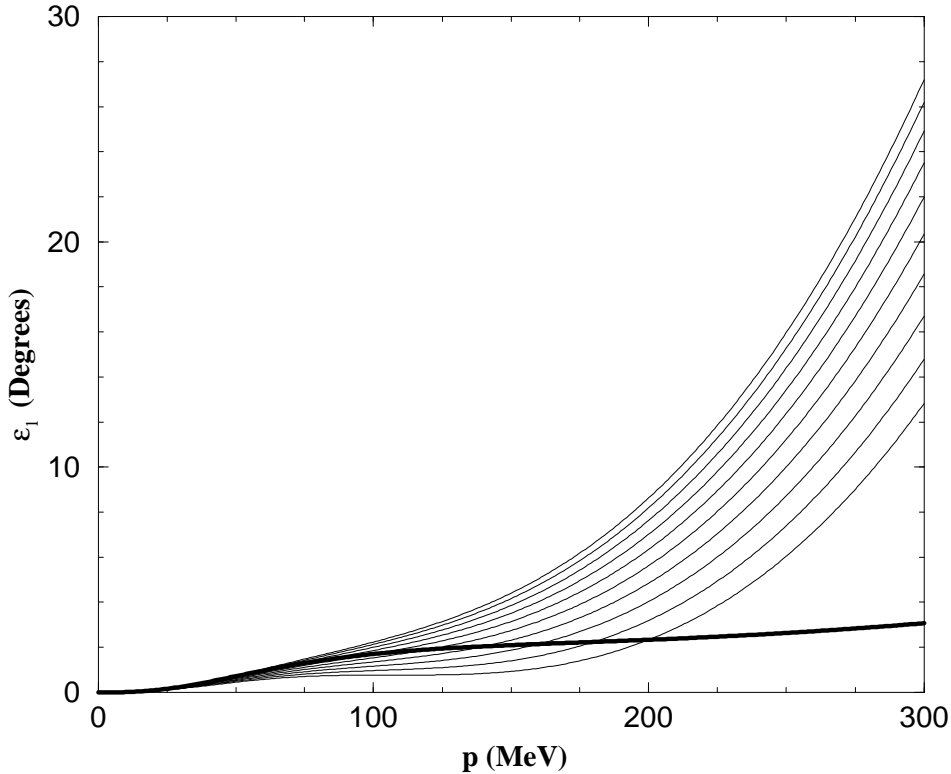}
\caption{Results of the variation of the renormalization condition.  The bold line is the multi-energy Nijmegan partial wave analysis.  The thin lines are the predictions resulting from fits at $p_{\sc r}\, = \,20 - 200$ MeV, in $20$ MeV intervals.}
\label{graph1}
\end{figure}

Assuming even modest convergence of the $Q$ expansion ($\Lambda \gtrsim 300$MeV), it should be reasonable to fit $\zeta_6$ at any $p_{\sc r} < 200$ MeV as we are working at NNLO.  By varying $p_{\sc r}$ from $0 - 200$MeV, a continuous set of presumably valid renormalization conditions is obtained which can be used to estimate the size of theoretical uncertainties present at this order due to higher-order terms.  In this manner, the range resulting from the variation of the $p_{\sc R}$ can be used as a gauge for the predictive power of the theory.  The results of this variation are displayed in Fig. \ref{graph1}.
Figure \ref{graph1} is an indication that the $Q$ expansion is not very predictive at this order.  By $100$ MeV, $100\%$ differences are present between the fits at $20$ MeV and $200$ MeV.  By $200$ MeV the differences reach $200\%$.

The great sensitivity to the choice  of renormalization condition casts doubt
on the usefulness of perturbative pions at least for this quantity at
 this order.  Another indication of a general problem with the approach is
the overall shape of the curve.
It is easy to see that for all $p \gg p_{\sc r}$, the EFT prediction overshoots the data as it climbs essentially quadratically, failing entirely to predict the slow linear ascent of the data.  The difficulty in reproducing the detailed shape is not immediately apparent in the work of FMS.   By considering only comparatively small momenta (150 MeV or less)
and choosing a relatively high renormalization condition, $p_{\sc r} \approx 110-120$ MeV, FMS have left very little room for this serious problem to manifest itself.  Note, however,  it is inconsistent with the power counting rules of $Q$
counting to restrict $p_{\sc r}$ to momenta below 150 MeV.  The counting rules were derived for $p \sim m_\pi$.  It is extremely implausible that momenta just above 150 MeV
cannot be considered of order $m_\pi$.

\section{Cancelations at NNLO}

The great sensitivity to the choice of renormalization condition is in some sense
paradoxical.  The values of  $\zeta_6$  for our fits which cross the data vary only from .359 (for $p_R$ of 20 MeV) to .412 (for $p_R$ of 200 MeV).
The apparent paradox is in explaining how 
 such modest variations in the
 fit parameter can yield such large variations in the predicted 
values.  This is particularly 
hard to understand in light of the fact that the contribution due to $\zeta_6$
is subleading and should have modest effects to the total, regardless of its value.  The resolution is both quite simple and very troubling:
there are cancelations between terms entering at NNLO which
are  much larger than the total value $\epsilon_1$ itself. 
 These cancelations can be seen by breaking the  NNLO contribution, 
$\epsilon_1^{(2)}$, into two pieces: the term proportional to $\zeta_6$, which we denote by $\epsilon_{1 ({\sc CT})}^{(2)}$, where the subscript CT indicates counterterm , and the remaining piece which we will denote as $\tilde{\epsilon}_1^{(2)}$.
The expression  $\epsilon_{1 ({\sc CT})}^{(2)}$ is :
\begin{equation} 
 \epsilon_{1 ({\sc CT})}^{(2)} \, = \,
- \frac{\zeta_6}{m_{\pi}^2} \frac{p^3}{\sqrt{\gamma^2 + p^2}}
\label{fitterm}
\end{equation}
The expression for $\tilde{\epsilon}_1^{(2)}$ can be extracted from Eqs.~(\ref{eps2}),(\ref{chi}), and (\ref{asd1}) with 
$\zeta_6$ set equal to zero. 

\begin{figure}
\epsfig{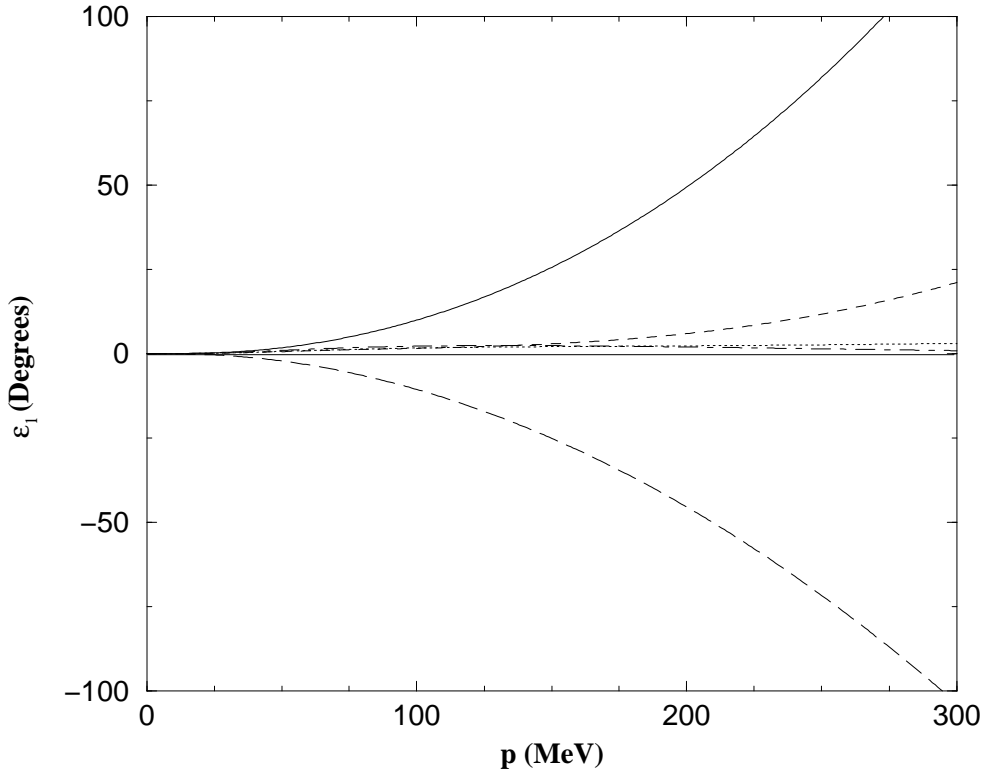}
\caption{$\epsilon_1$ at NNLO.  The NNLO contribution to $\epsilon_1$ is divided into two pieces,  $\tilde{\epsilon}_1^{(2)}$ (solid line) and $\epsilon_{1 ({\sc CT})}^{(2)}$ (long dashed line).  The multi-energy Nijmegan partial wave analysis  is the dotted line.  The NLO and the full NNLO calculations are graphed as the short-dashed, and dot-dashed lines, respectively. }
\label{graph2}
\end{figure}

In Fig. \ref{graph2} we show the $\epsilon_1$ mixing angle extracted from the
Nijmegen PWA, along with the NLO prediction, the full NNLO calculation of $\epsilon_1$
(with $\zeta_6$ fixed using $p_{\sc R} = 100$ MeV ),  and the two terms which enter at
NNLO---$\epsilon_{1 ({\sc CT})}^{(2)}$ and  $\tilde{\epsilon}_1^{(2)}$.  The
$\tilde{\epsilon}_1^{(2)}$ term and the $\epsilon_{1 ({\sc CT})}^{(2)}$ term are 
typically more than an order of magnitude  larger than the prediction at NLO, the
prediction at NNLO, and the result extracted from scattering data.  Both of these
terms are much larger than the leading order terms suggesting serious problems with
convergence.  Perhaps more seriously, they are much larger than the overall scale of the
final result.  Note  that formally both of these terms are subleading and should each
separately give modest corrections to the NLO result.  Instead, each swamps the NLO, and
our prediction depends on delicate cancelations between the NNLO terms.  The great
sensitivity to the choice of renormalization condition is a result of this requirement of
very large cancelations.  Rather slight changes in the momentum where one fits
translate into rather large changes in the sum.  
 
The large scale of the cancelations at NNLO suggest that there is no real
predictive power at this order.  The individual quantities which cancel
should have corrections of relative order $Q/\Lambda$ due to higher-order
contributions; such corrections can easily be $30\%$. Since these
terms are required to cancel by more than an order of magnitude to reach
the data, $\sim 30\%$ uncertainties in each term translate to $\sim 300 \%$
uncertainties for the total.   Thus, there appears to be no predictive
power at this order.  Indeed, this analysis based on the cancelations
suggests that lack of predictive power is more severe than one might have
concluded on the basis of the sensitivity to the renormalization condition
found in the previous section.

One practical consquence of the large cancelations at NNLO and the resultant
great sensitivity to the value of the parameters is evident in the calculation
seen in Fig.~(\ref{FMSfig}).  Using the parameters quoted in Ref. \cite{FMS} we produced
a plot which differed  from the plot in Ref. \cite{FMS} by a perceptable amount.  
We believe this small difference was a consquence of extremely small rounding errors
whose effects were magnified.  By
adjusting  the $\zeta_6$ parameter down from .385 to .3835 we were able to qualitatively
reproduce their result.

Of course, it is logically possible that the large cancelation seen at NNLO is
an artifact of the  way the NNLO contribution was divided into two terms.  One
can always divide a small number into the difference of two unnaturally large
numbers; in doing this  one can only conclude that there is something wrong
with the way the contribution was divided, and not that there is any
problem with underlying theory.   However, in the present context this appears
to be quite unlikely  {\it a priori}.  The division into these two terms was
not chosen to maximize their individual size, but rather to see the size of the
counterterm contribution (which by standard power counting arguments
contributes at NNLO and should have the characteristic size of a typical NNLO
contribution).  Ultimately what we see is that the scale of the NNLO
contribution is much larger than the full result and one is depending entirely
on the fitting $\zeta_6$ to  force the result to be at more or less the
right size.
 
As noted above, it is surprising that the sensitivity of the NNLO prediction
of $\epsilon_1$ to the choice of renormalization condition, although large,
 is as small as
it is given the very large cancelations at NNLO.  This can be understood,
in part, in terms of the detailed shapes of the two terms at NNLO. 
Both the $\epsilon_{1 ({\sc CT})}^{(2)}$ and $\tilde{\epsilon_1}^{(2)}$ terms
are  cubic in $p$ for low $p$, and essentially quadratic for $p \gtrsim
100$ MeV.  Thus the shapes are rather similar; if they largely cancel at
one momentum they will continue to largely cancel at other momenta.
This tends to moderate the sensitivity to the choice of renormalization condition.

The $\epsilon_1$ extracted from the scattering 
data  is also  cubic in $p$ at very low momenta, and with one free
parameter, it is not surprising that the theory fares moderately well in
this regime.  However, when fits are performed at $p_{\sc r} > 100$ MeV, the
quality of the fits worsen dramatically - a result of trying to fit an
essentially quadratic function to a linear one.  As the momentum increases
, this quadratic function dominates since the cancelation between   $\epsilon_{1 ({\sc CT})}^{(2)}$ and $\tilde{\epsilon_1}^{(2)}$ is not perfect.
Therefore it seems
reasonable to conclude that the theory is simply not predictive above
$100$ MeV.  

It is amusing to note that at  extremely low momenta $\ll 50$ MeV, all of the fits done  with $p_{\sc R}$ as large as 200 MeV seem to converge to a cubic with approximately the same coefficient at low momentum. 
 Given the fact that theory does not appear to be predictive at 200 MeV, we are forced to conclude that this is fortutitous.

\section{Discussion}

In the Introduction we raised the question of whether the chiral
part of $Q$ counting was converging well.  In this paper we have
seen that $\epsilon_1$ at NNLO does not appear to be predictive
and hence does not support the assertion that perturbative pions
are useful.  It is useful, however, to explore the question of
whether one would expect the $Q$ counting expansion to be predictive
at this order.  In fact, even if $Q$ counting were converging
one might have suspected on the basis of simple power counting
arguments that the S-D mixing angle would not be
predictive at NNLO
without  actually doing the calculation. 

 The leading nonvanishing order contribution to $\epsilon_1$ is order $(Q/\Lambda)^1$. 
To estimate the expected size of $\epsilon_1$ one needs a characteristic value of $Q$
and $\Lambda$.  It is not immediately clear what values one should take for $Q$ or
$\Lambda$.  It is probably sensible to consider $\epsilon_1$ at a momentum of $\sim
m_\pi$ so that one can take $Q \sim m_\pi$. If $\Lambda \sim 300$ MeV as suggested by
Kaplan, Savage and Wise \cite{KSW} the power counting estimate for $\epsilon_1$ is $\sim
25^{\circ}$, which is more than an order magnitude larger than the observed value of $\sim
2^{\circ}$.  Assuming that the $Q$ expansion is sensible, this suggests that
$\epsilon_1$ is itself a result of large accidental cancelations.  As an unnaturally
small observable, theoretical difficulties are to be expected, and  expansions of
$\epsilon_1$ are expected to converge very slowly if at all.  For example, corrections
to the NNLO are nominally $(Q/\Lambda)^3$ corresponding to $\sim 6^{\circ}$ which is
three times the observed value, indicating a complete lack of predictive power at this
order.  

If $Q$ counting is in fact converging, but $\epsilon_1$ is just accidently small, one can
estimate to what order one must work in order to get an accurate prediction.  If one
insists on an accuracy of $\sim 20\%$, one estimates contributions at their nominal
order  and $\Lambda$ is taken to be 300 MeV, then one has to work to order
$(Q/\Lambda)^7$, this corresponds to
next-to-next-to-next-to-next-to-next-to-next-to-next-to-next-to leading order.  Thus,
even if the theory is converging as well as could be expected for $\Lambda \sim $ 300
MeV, it is implausible that such calculations will ever prove tractable.  Of course if 
$\Lambda$ is larger, one need not work to quite as high an order.  Nevertheless, if
$\Lambda \sim 3 m_\pi$ one needs to work to order  $(Q/\Lambda)^5$ which also seems to
be intractable. 

Naively it seemed that the s-d mixing parameter, $\epsilon_1$, should be a good
place to test the chiral part of the $Q$ expansion, as it is chirally sensitive 
in the sense defined in the Introduction.  However, the unnaturally small size
of the mixing vitiates its utility as a strong  test of perturbative pions.
We do note, though, that the  very fact that the mixing is unnatural in itself
does cast some doubt on the use of perturbative pions in general, as unnatural
results are, by definition, unexpected if the scales in the problem are properly 
identified.

In summary, we
find that calculations of $\epsilon_1$ at NNLO do not appear to be predictive
as they are highly dependent on the fitting procedure used ({\it i.e.}, on the renormalization condition) and the fitting involves delicate cancelations of quantities much larger than the total value.  We argue that this lack of
predictive power is consistent with simple power counting estimates given
the small size of $\epsilon_1$.  Thus, we find that calculations of $\epsilon_1$
to this order do not provide evidence for the use of perturbative pions.

\end{document}